\documentclass[twocolumn,showpacs,preprintnumbers,amsmath,amssymb]{revtex4}
\usepackage{graphicx}

\begin{document}


\title{Topological Symmetry And Existence of Partial Synchronization}

\author{Bin Ao$^{1,\ 2}$}
\author{Xin Qi$^{1}$}
\author{Zhi-Gang Shao$^{1,\ 2}$}
\author{Lei Yang$^{1,\ 3}$\footnote{Corresponding author: lyang@impcas.ac.cn}}

\affiliation{$^{1}$Institute of Modern Physics, Chinese Academy of
Science, Lanzhou 730000, China}

\affiliation{$^{2}$Department of Physics, Centre for Nonlinear
Studies, and The Beijing-Hong Kong-Singapore Joint Centre for
Nonlinear and Complex Systems (Hong Kong), Hong Kong Baptist
University, Kowloon Tong, Hong Kong, China}

\affiliation{$^{3}$Department of Physics, Lanzhou University,
Lanzhou 730000, China}

\date{\today}

\begin{abstract}
We study the relationship between the partial synchronous (PaS)
state and the coupling structure in general dynamical systems. By
the exact proof, we find the sufficient and necessary condition of
the existence of PaS state for the coupling structure. Our result
shows that the symmetry of the coupling structure is not the
equivalent condition which is supposed before but only the
sufficient condition. Furthermore, for the existence of the PaS
state, the general structure is the equal-degree random.
\end{abstract}

\maketitle

An interesting form of dynamical behavior occurs in the coupled
systems when only some of the subsystems behave in the same way,
which is Partial Synchronization (PaS). The behavior has attracted
much attention in the physical \cite{application_phy}, biological
\cite{application_bio}, ecological \cite{application_eco}, and
other systems. As the coupling strength between the subsystems is
small enough, the state of coupled system is turbulence.
Increasing the coupling strength, the PaS state is often observed
before the Global Complete Synchronization (CS). The existence of
various PaS states leads to the remarkably complex dynamical
behaviors \cite{global} in the coupled systems.

\begin{figure}[tbp]
\includegraphics[width=1.5in,height=1.3in]{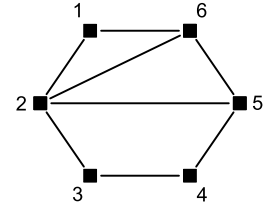}
\caption{A topological structure of the coupled system without any
partial synchronous solution.} \label{nonexistence}
\end{figure}

Considering the dynamics \cite{sys}
\begin{eqnarray}
\textbf{X}_{n+1}=\textbf{F}(\textbf{X}_n)+\varepsilon\Gamma\otimes
C \textbf{F}(\textbf{X}_n), \label{map}
\end{eqnarray}
where $\mathbf{X}=(\mathbf{x}^{1},\mathbf{x}^{2},\cdots
,\mathbf{x}^{N})$ represents the status of the system. The
independent state, which is defined as $\mathbf{x}^{1}\neq
\mathbf{x}^{2}\neq \cdots \neq \mathbf{x}^{N}$, and the CS state,
which is defined as $\mathbf{x}^{1}=\mathbf{x}^{2}=\cdots
=\mathbf{x}^{N}$, exist for common coupling ways \cite{couple}
between the subsystems. However it is possible for a given coupled
system without any PaS solutions, Fig. \ref{nonexistence} just
shows an example. Substituting all possible 365 PaS solutions
(e.g., $\mathbf{x}^1=\mathbf{x}^2$ and
$\mathbf{x}^3\neq\mathbf{x}^4\neq\mathbf{x}^5\neq\mathbf{x}^6$)
into Eq. (\ref{map}) at the coupling structure shown in Fig.
\ref{nonexistence} will give false statements.

In recent years, it is believed there are some tight relations
between the topological symmetry of the coupling structure and the
PaS state. For example, the asymmetric PaS pattern that disagrees
with a symmetrical structure has never been observed
\cite{symm_manrubia}; the symmetry group theory can be used to
describe the partial periodical synchronous state in some regular
structures with the same symmetry \cite{symm_yu}; all of PaS
states corresponding to each topological symmetry in a ring were
observed \cite{symm_zhang}, etc. So one could suppose that the
symmetry is the necessary and sufficient condition for the
existence of the PaS state.

\begin{figure}[tbp]
\includegraphics[width=3.5in,height=1.7in]{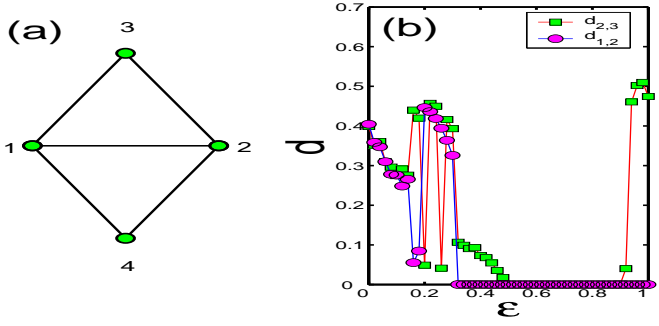}
\caption{(a) A simple, but typical, network with symmetries. (b)
The average distance $d_{1,2}$ and $d_{2,3}$ versus the coupling
strength $\varepsilon $.} \label{symm_example_n4}
\end{figure}

For small amount of subsystems, it seems that the above
relationship comes into existence. The coupled structure Fig.
\ref{symm_example_n4}(a) can be represented mathematically by the
adjacent matrix
\begin{equation*}
\begin{array}{cc}
A_{4}=\left(
\begin{array}{cccc}
0 & 1 & 1 & 1 \\
1 & 0 & 1 & 1 \\
1 & 1 & 0 & 0 \\
1 & 1 & 0 & 0%
\end{array}%
\right). &
\end{array}%
\end{equation*}%
Here, node 1 and 2 is symmetric in the structure, and $A_4$ will
be invariable under a permutation transformation $1\rightarrow2$
or $2\rightarrow1$. The curves $d_{1,2}(\varepsilon)$,
$d_{2,3}(\varepsilon)$ \cite{aver_dis} are shown in Fig.
\ref{symm_example_n4} (b), $\varepsilon$ is the coupling strength.
The synchronous solution
$\mathbf{x}_1=\mathbf{x}_2\neq\mathbf{x}_3\neq\mathbf{x}_4$ is
observed in the region $[0.3,0.45]\cup[0.7,1]$ of $\varepsilon$.
Thus, the PaS state will be achieved with the corresponding
symmetry in $A_4$ among the symmetrical nodes.

\begin{figure}[tbp]
\includegraphics[width=3.5in,height=1.7in]{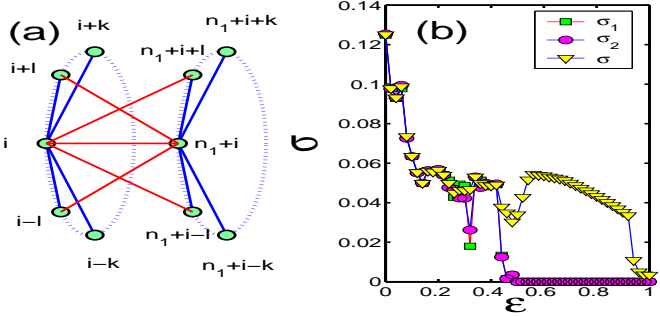}
\caption{(a) The scheme of a topological structure with the
"rotate" symmetry. (b) The variation of the two clusters
$\sigma_1$, $\sigma_2$ and the whole system $\sigma$ as functions
of $\varepsilon$ with $n_{1}=n_{2}=100$, $k=40$, $l=10$.}
\label{symm_example_n100}
\end{figure}

A more complex case is shown in Fig. \ref{symm_example_n100} (a)
with the same dynamics as above. There are two clusters and their
nodes are denoted by $ 1,2,\cdots ,n_{1}$ and
$n_{1}+1,n_{1}+2,\cdots, n_1+n_2$. Node $i$ $(1\leq i\leq n_1)$ is
coupled to the nodes $i-k, i-k+1, \cdots, i+k$, $ n_1+i-l,
n_1+i-l+1, \cdots, n_1+i+l$, and node $n_1+i$ $(1\leq i\leq n_1)$
is coupled to the nodes $n_1+i-k$, $n_1+i-k+1, \cdots$, $n_1+i+k$,
$i-l, i-l+1, \cdots, i+l$. Obviously, there is "rotate" symmetry
in every cluster. The adjacent matrix will be invariant under a
"rotate" permutation transformation in each of them: e.g., it is
$1\rightarrow 2$, $2\rightarrow 3$, $\cdots$, $n_{1}-1\rightarrow
n_{1}$, $n_{1}\rightarrow 1$. We define a $M\times M$ matrix
$R_{M}$, where $(R_{M})_{M,1}=1$, $(R_{M})_{i,i+1}=1$
$(i=1,2,\cdots ,M-1)$, and other elements of $R_{M}$ are 0. Thus
the permutation matrix \cite{perm_matr} of this transformation
will be $T_{d}=R_{n_{1}}\oplus R_{n_{2}}$, where "$\oplus$" is
direct sum of two matrices. Fig. \ref{symm_example_n100} (b) shows
the time average of the variation of all subsystems
$\sigma(\varepsilon)=\lim_{N'\to\infty}\frac{1}{N'}\sum_{n=1}^{N'}\sqrt{\sum_{i=1}^N(\mathbf{x}^i_n-\frac{1}{N}\sum_{i=1}^N\mathbf{x}^i_n)^2}$,
and of the two clusters $\sigma_1(\varepsilon)$,
$\sigma_2(\varepsilon)$ with $n_{1}=$ $n_{2}=100$, $k=40$, $l=10$.
For $\varepsilon\in[0.45,1]$, $\sigma_1=\sigma_2=0$, namely
$\mathbf{x}_1=\cdots=\mathbf{x}_{n_1}$,
$\mathbf{x}_{n_1+1}=\cdots=\mathbf{x}_{N}$. The PaS solution of
the "rotate" symmetrical nodes is observed.

In this Letter, we investigate in detail the relationships between
the PaS solution and the coupling structure. The PaS solution is
defined as follow: for a dynamical system with phase space
$\mathbb{R}^{Nm}$, a $K$-cluster synchronous solution is a
$K$-dimensional subspace, denoted by $V$, of $\mathbb{R}^{Nm}$. It
can be represented by
\begin{equation}
\begin{array}{c}
\mathbf{x}^{i_{1}^{1}}=\mathbf{x}^{i_{1}^{2}}=\cdots =\mathbf{x}%
^{i_{1}^{n_{1}}} \\
\mathbf{x}^{i_{2}^{1}}=\mathbf{x}^{i_{2}^{2}}=\cdots =\mathbf{x}%
^{i_{2}^{n_{2}}} \\
\cdots \cdots \\
\mathbf{x}^{i_{K}^{1}}=\mathbf{x}^{i_{K}^{2}}=\cdots =\mathbf{x}%
^{i_{K}^{n_{K}}}%
\end{array}
\label{syn_sta_1}
\end{equation}
where $\mathbf{x}^{i_{a}^{b}}$ denotes the $b$th subsystem in the
$a$th cluster and $\{n_{i}\}_{i=1}^{K}$ is the size of each
cluster that satisfied $\sum_{i=1}^{K}n_{i}=N$ ($N>K$ $>1$). The
CS and independent solution are the particular cases of the
definition for $K=1$ and $K=N$. The relationship between PaS
solution and the coupling structure could be described by two
questions as follows:

Question A: If one finds symmetry in a coupling structure, can a
corresponding PaS solution be obtained?

Considering the matrix form of Eq. (\ref{syn_sta_1}),
\begin{equation}
T\mathbf{X}=\mathbf{X},\forall \mathbf{X}\in V,  \label{syn_sta}
\end{equation}
where $T$ is a permutation matrix. $\mathbf{X}\in V$ is invariant
under the transformation $T$, so $V$ is the invariant subspace of
$T$ and the eig-subspace of $T$ corresponding to eigvalue 1. That
$V$ is the invariant subspace of the dynamical system Eq.
(\ref{map}) requires
\begin{equation}
C\mathbf{X}\in V,\ \ \ \forall \mathbf{X}\in V. \label{solution}
\end{equation}
Next, if there is a symmetry $T$ in structure $C$, then $C$ will
be invariant under a permutation transformation $T$. The
mathematical representation is
\begin{equation}
T^{-1}CT=C,  \label{symm_net}
\end{equation}
Therefore, the matrices $C$ and $T$ are commute, or
\begin{equation}
CT=TC.  \label{commute}
\end{equation}

Question A can be represented by a mathematical statement as
follow:
$$
\begin{array}{cc}
If & T\mathbf{X}=\mathbf{X},\\
then & Eq.\ (\ref{symm_net}) \Rightarrow Eq.\ (\ref{solution})\\
\end{array}
$$
Multiplying the two sides of Eq. (\ref{commute}) by $\mathbf{X}\in
V$, we have
\begin{equation}
TC\mathbf{X}=CT\mathbf{X},\forall \mathbf{X}\in V;
\label{proof_suf_1}
\end{equation}
Combining Eq. (\ref{syn_sta}) with Eq. (\ref{proof_suf_1}) gives
\begin{equation}
TC\mathbf{X}=C\mathbf{X},\forall \mathbf{X}\in V.
\label{proof_suf_2}
\end{equation}
So, $C\mathbf{X}$ is also the eigvector of $T$ with eigvalue 1.
And then, Eq. (\ref{solution}) will be satisfied for all
$\mathbf{X}\in V$. Thus, we conclude that, for a symmetrical
structure, the dynamical system has a corresponding PaS solution.

Question B: If one finds a PaS solution, can the corresponding
symmetry in the coupling structure be observed.

\begin{figure}
\includegraphics[width=3.5in,height=1.7in]{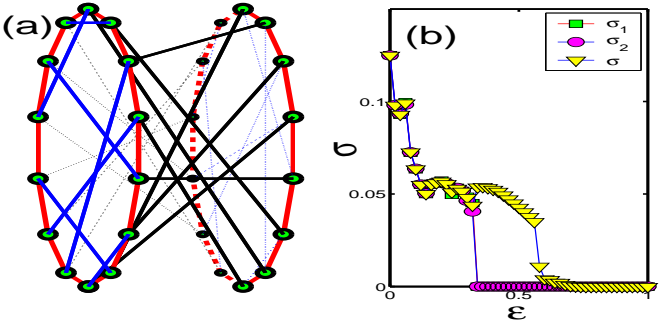}
\caption{(a) The scheme of an equal degree random structure. (b)
The variance of the two clusters $\protect\sigma _{1}$, $\protect
\sigma _{2}$ and the whole network $\protect\sigma $ as functions
of the coupling strength $\protect\varepsilon$ in the parameter
$n_{1}=n_{2}=100$, $p=0.5$, $p_{r}=1$. } \label{edr}
\end{figure}

Here, an interesting example is shown in Fig. \ref{edr} (a).
Considering two clusters, each one has $n$ subsystems, every
subsystem is randomly connected to $[pn]+1$ subsystems in the same
cluster ($p$ is a probability and $[pn]$ means the integer part of
$pn$) and $[p_{r}n]+1$ subsystems in another cluster ($p_{r}$ is
also a probability). Each subsystem in each cluster has equal
degree and the connections between two clusters are also
equal-degree. Fig. \ref{edr} (b) shows the variance of the two
clusters ($\sigma _{1}$, $\sigma _{2}$) and the whole system
($\sigma $) as functions of the coupling strength $\varepsilon $
in the parameters $n=100$, $p=1$,
$p_{r}=0.5$. As $\varepsilon \in \lbrack 0.34,0.72 \rbrack$, $\sigma _{1}=$$%
\sigma _{2}=0$ and $\sigma >0$, the PaS solution is observed. Due
to the random connections between the subsystems, there is not
symmetry in the structure.

Question B can be represented by a mathematical statement as
follow:
$$
\begin{array}{cc}
If & T\mathbf{X}=\mathbf{X},\\
then & Eq.\ (\ref{solution}) \Rightarrow Eq.\ (\ref{symm_net})\\
\end{array}
$$
The following relations could be derived
\begin{eqnarray}
&C\mathbf{X}\in V,T\mathbf{X}=\mathbf{X},\forall \mathbf{X}\in V  \notag \\
\Longleftrightarrow &TC\mathbf{X}=C\mathbf{X}  \notag \\
\Longleftrightarrow &TC\mathbf{X}=CT\mathbf{X}  \notag \\
\Longleftrightarrow &(T^{-1}CT-C)\mathbf{X}=0  \notag \\
or & (CT-TC)\mathbf{X}=0.  \label{solution_exist}
\end{eqnarray}
Obviously, Eq. (\ref{solution_exist}) is not equivalent to Eq.
(\ref{symm_net}). We can conclude that it is possible for a PaS
solution without any symmetry in a dynamical system. Fig.
\ref{edr} just gives an example. Another important feature of Eq.
(\ref{solution_exist}) is that, in fact, the sufficient and
necessary condition can be drawn from Eq. (\ref{solution_exist}).

The component form of Eq. (\ref{solution_exist}) is
\begin{equation}
\sum_{n=1}^{N}C_{mn}\mathbf{x}^{n}-\sum_{j=1}^{N}C_{ij}\mathbf{x}%
^{j}=0, \forall
\mathbf{X}=(\mathbf{x}^{1},\mathbf{x}^{2},\cdots,\mathbf{x}^{N})\in
V. \label{solution_exist_comp}
\end{equation}
for $T_{im}=1$. Relabelling the subsystems in order to collect
together the subsystems in the same cluster, Eq. (\ref{syn_sta_1})
will be rewritten in the form
\begin{equation}
\begin{array}{c}
Cluster 1:\ \\
\mathbf{x}^{1}=\mathbf{x}^{2}=\cdots
=\mathbf{x}^{n_{1}}\equiv \mathbf{y}^{1}\\
\cdots\ \cdots \\
Cluster s:\ \\
\mathbf{x}^{\sum_{k=1}^{s-1}n_{k}+1}=\mathbf{x}^{\sum_{k=1}^{s-1}n_{k}+2}=\cdots =\mathbf{x}^{\sum_{k=1}^{s-1}n_{k}+{n_{s}}%
}\equiv \mathbf{y}^{s} \\
\cdots\ \cdots \\
Cluster K:\ \\
\mathbf{x}^{\sum_{k=1}^{K-1}n_{k}+1}=\mathbf{x}^{\sum_{k=1}^{K-1}n_{k}+2}=%
\cdots =\mathbf{x}^{\sum_{k=1}^{K-1}n_{k}+n_{K}}\equiv
\mathbf{y}^{K}.
\end{array}
\label{syn_sta_2}
\end{equation}
And the general form \cite{t4td} of $T$ will be
\begin{equation}
T=\bigoplus_{i=1}^{K}R_{n_{i}},i=1,2,\cdots ,K. \label{T_form}
\end{equation}
Then the general form of $\mathbf{X}$ will be
$\mathbf{X}=(\mathbf{y}
^{1}1_{1,n_{1}},\mathbf{y}^{2}1_{1,n_{2}},\cdots
,\mathbf{y}^{K}1_{1,n_{K}})^T$ where $1_{M,N}$ is an $M\times N$
matrix in which every element is 1. Eq.
(\ref{solution_exist_comp}) is the $i$th row of Eq.
(\ref{solution_exist}) and $\{y^{s}\}_{s=1}^{K}$ are independent,
thus we can collect Eq. (\ref{solution_exist_comp}) into $K$ terms
and the $s$th ($s=1,2,\cdots,K$) term describes the degree of
subsystem $i$ contributed by $s$th cluster. If subsystem $i$
belongs to cluster $s'$, then the $s$th term will be
\begin{equation}
\mathbf{y}^{s}\sum_{j=N_{s}+1}^{N_{s}+n_{s}}(C_{ij}-C_{mj})=0,(i=N_{s^{\prime}}+1,N_{s^{\prime}}+2,\cdots
,N_{s^{\prime }}+n_{s^{\prime}})\label{equ_deg}
\end{equation}
where $N_{s}=\sum_{i=1}^{s-1}n_{i}$ and $m=i+1$ when $i<N_{s^{\prime }}+n_{s^{\prime }}$; $%
m=1$ when $i=N_{s^{\prime }}+n_{s^{\prime }}$.

Considering two different situations in Eq. (\ref{equ_deg}), we
can draw two statements as follow:

\textsl{\textbf{S1}}: $s\neq s^{\prime }$.
$\sum_{j=N_{s}+1}^{N_{s}+n_{s}}C_{ij}$ is the degree of subsystem
$i$, which is contributed by the $s$th cluster. So Eq.
(\ref{equ_deg}) shows that \textbf{the degrees of subsystems in a
cluster} (e.g., the $s^{\prime }$th cluster), \textbf{which are
contributed by the subsystems in another cluster} (e.g., the $s$th
cluster), \textbf{should be the same.}

\textsl{\textbf{S2}}: $s= s^{\prime }$. Since $C_{mm}=C_{ii}=-1$,
the rest part of Eq. (\ref{equ_deg}) illustrates that \textbf{the
subsystems' degrees contributed by their cluster also should be
the same}.

The two statements are the complete representation of Eq.
(\ref{solution_exist}) and now we have a clear physical picture of
the necessary and sufficient condition of the existence of a PaS
solution. Then, by using the two statements in the particular
cases, we can obtain some interesting results:

\textbf{I}. The nonexistence of any PaS states in the system shown
in Fig. \ref{nonexistence} is easy to be proved. Now we do not
need to substitute all the possible solutions into the dynamical
system. One can suppose that there is at least one PaS solution.
According to \textbf{S1} and \textbf{S2}, the PaS should be
observed between the subsystems with the same degrees, so the
subsystem 2 itself must be a PaS cluster without anyone else. On
the other hand, subsystem 2 is connected to subsystems 1, 3, 5, 6.
\textbf{S1} requires that these four subsystems should be included
in one or more clusters, but any of these clusters will not
include subsystem 4, because the degree of subsystem 4 contributed
by node 1 is 0 while that of others is 1, i.e., subsystem 4 itself
should be a cluster too. And then, subsystem 4 is connected to
subsystem 3 and 5, and because of the same reason, subsystem 3, 5
should be in different clusters with subsystem 1, 6. However, the
degrees of subsystem 3 and subsystem 1 are different with that of
subsystem 5 and subsystem 6 respectively, that means they also
should be in different clusters. So, we can see there will be at
least 6 clusters in this system with just 6 subsystems, any PaS
phenomenon will not be observed in this system.

\textbf{II}. If there is symmetry in a dynamical system, Eq.
(\ref{symm_net}) can be rewritten as
\begin{equation}
C_{ij}=C_{mn},\label{symm_comp}
\end{equation}
where $i=N_{s^{\prime }}+1,$ $N_{s^{\prime }}+2,\cdots,$
$N_{s^{\prime }}+n_{s^{\prime}}$; $j=N_s+1,$ $N_s+2,\cdots,$
$N_s+n_s$, $N_{s}=\sum_{i=1}^{s-1}n_{i}$ and $m=i+1$ when
$i<N_{s^{\prime }}+n_{s^{\prime }}$; $m=1$ when $i=N_{s^{\prime
}}+n_{s^{\prime }}$; $n=j+1$ when $j<N_{s^{\prime }}+n_{s^{\prime
}}$; $n=1$ when $j=N_{s^{\prime }}+n_{s^{\prime }}$. So Eq.
(\ref{symm_comp}) is a stronger condition than Eq.
(\ref{equ_deg}). That is why in few body systems
\cite{symm_manrubia, symm_yu, symm_zhang} there are tight
relations between the PaS solution and the symmetry in the
coupling structure. Furthermore, one can prove that if there are
only one or two subsystems in every cluster
($n_{s^{\prime}},n_s=1$ or $2$, $s,s'=1,2,\cdots,K$) then there
should be at least one symmetry in the coupling structure
($n_{s^{\prime}},n_s=2$ is nontrivial). In this case, the
$(C_{ij})_{n_{s^{\prime}}\times n_s}$ can only be
$\left(\begin{array}{cc}
0 & 1 \\
1 & 0 \\
\end{array}\right)$
or $\left(\begin{array}{cc}
1 & 1 \\
1 & 1 \\
\end{array}\right)$
and Eq. (\ref{symm_comp}) will be satisfied automatically.

\textbf{III}. The CS, as a special case of the PaS, can also be
included in the Eq. (\ref{solution_exist}), where the dimension of
$V$ is only 1. And its normalized basis vector is
$\mathbf{X}_{global}=1_{N,1}/\sqrt{N}$. In fact, this vector is
always the eigvector of all of $T$.
because every PaS subspace contains the CS subspace ($\mathbf{x}^{1}=\mathbf{%
x}^{2}=\cdots =\mathbf{x}^{N}$). The $C$ is a non-row matrix (the
summery of any row is 0) and the matrix product $TC$, $CT$ also
are. Thus, we conclude that, for the coupled system Eq.
(\ref{map}), the CS solution always exists no matter what the
coupling structure $C$ is. So Eq. (\ref{solution_exist}) can be
satisfied for all of $C$ in this case, i.e., the CS solution
exists in every system Eq. (\ref{map}) no matter what the system
topology $C$ is.

In conclusion, we have studied the relationship between the
coupling structure and the PaS state in general dynamical systems.
The sufficient and necessary condition of the existence of PaS
state for the coupling structure is found by the exact proof. And
the result is counterintuitive, that is, the existence of PaS
state doesn't require the symmetry in the coupling structure,
which is supposed before. Then, a new structure, the equal-degree
random structure, is obtained. According to the sufficient and
necessary condition, it is the general structure for the existence
of the PaS state. Finally, we should stress that the proof also
can be applied to the differential dynamical systems and the
conclusions are the same.

The authors are grateful for discussions with Ping Lin and Jinfang
Zhang. The work was in part supported by the 100 Person Project of
the Chinese Academy of Sciences, the China National Natural
Science Foundation with Grant No. 10775157, the Hong Kong Research
Grants Council (RGC), and Hong Kong Baptist University.


\begin{thebibliography}{99}
\bibitem{sys} This is one of widely used forms of the coupling nonlinear maps.
One can apply certain $m$-dimensional map $\mathbf{x}_{n+1}=
\mathbf{f}(\mathbf{x}_{n})$,
$\mathbf{x}=(x_{n}^{1},x_{n}^{2},\cdots ,x_{n}^{m}) $ as subsystem
to construct a coupled system
$\mathbf{X}=(\mathbf{x}^{1},\mathbf{x}^{2},\cdots
,\mathbf{x}^{N})$,
$\mathbf{F}=(\mathbf{f}^{1},\mathbf{f}^{2},\cdots
,\mathbf{f}^{N})$. The common interesting is the linear coupling
between the subsytems, thus the coupling term will be $\varepsilon
\Gamma \otimes C\mathbf{F}(\mathbf{X})$, where $\varepsilon $ the
coupling strength. $\Gamma :\mathbb{R}^{m}\rightarrow
\mathbb{R}^{m}$ characterizes the coupling scheme among the
subsystems and $C$ is the Laplacian matrix. $C_{ij}=A_{ij}/k_{i}$
for $j\neq i$ and $C_{ii}=-1$, where $k_{i}$ is the degree of
subsystem $i$ and $A_{ij}$ is an element of the adjacent matrix
$A$ of the coupling structure. To gain insight in numerical
simulation, we set $f(x)$ to be the logistic map $f(x)=4x(1-x)$,
but the proof do not depend on the dynamics.
\bibitem{aver_dis} In this system, one can define the average distance
between two oscillators $i$ and $j$ by $d_{ij}=\lim_{N\rightarrow
\infty }(N^{-1}\sum_{n=1}^{N}\Vert
\mathbf{x}_{n}^{i}-\mathbf{x}_{n}^{j}\Vert )$, and say they are in
a synchronous state in the sense that $d_{ij}<10^{-12}$. And the
partial synchronization of a network means that some of nodes are
in a synchronous state while others not.
\bibitem{t4td}The $T_{d}$ in the second example is only a special case of
Eq.(\ref{T_form}), and the permutation transformation done to
$A_4$ can also be represented by $T_{4}=R_{2}\oplus 1\oplus 1$.
\end{thebibliography}
\end{document}